\documentclass[preprint2]{aastex}
\usepackage{natbib}

\def\kms{km s${}^{-1}$}

\begin{document}

\title{Australia Telescope Compact Array Observations of the OH Star Roberts
22: Resolved Images of OH Emission}

\author{K.K. Dyer} 

\author{W.M. Goss, A.J. Kemball} 




\begin{abstract}

We have imaged the unusual OH/IR object G 284-0.8 (Roberts~22) in the
OH transitions at 1612, 1665 and 1667 MHz using the Australia
Telescope Compact Array in Narrabri, NSW, Australia. 
The angular resolution of the OH images we present here is
6\arcsec \ (12000 AU at a distance of 2 kpc). We find OH emission, of
extent 1\farcs5\ (2800 AU), located inside the optical bipolar flow
region (size 8\arcsec) discovered by \citet{allen1980} and imaged
recently by \citet{sahai1999} with the Hubble Space Telescope. We
present astrometry of Roberts~22, providing registration of the OH and
Hubble Space Telescope images to within $\sim$0\farcs35. We describe the OH
velocity field in Roberts~22 and discuss the kinematic properties of the
source and its evolutionary status.

\end{abstract}

\section{Introduction}
\label{sec:introduction}

The OH maser source G 284.2-0.8 was discovered in the course of a
southern galactic plane survey at 1612 MHz near the \ion{H}{2} region RCW 48
(Manchester, Goss and Robinson 1969; Manchester, Robinson and Goss
1970). The discovery pre-dated the detection of OH emission from IR
stars, and the nature of the source was not appreciated at the time.
Subsequently it became clear that G 284.2-0.8 could be classified as a
Type II OH/IR star, as defined by \citet{wilson1970}. The source
acquired its most common name, Roberts~22 (hereafter R22), by its
inclusion in a catalog of Wolf-Rayet stars by
\citet{roberts1962}. Additional designations include Hen 404, Wray 549
and IRAS 10197-5750.

Subsequent single-dish observations using the Parkes radio telescope
from 1968 to 1978 were reported by \citet{allen1980}. These
observations found strong OH emission at 1612 MHz and 1665 MHz in a
double-peaked spectrum, with fainter emission at 1667 MHz, and no
detected emission in the 1720 MHz satellite line. The source was found
to have low levels of circular polarization. \citet{allen1980}
reported no time variations exceeding a factor of two over
this interval, and concluded that the maser is pump-limited. An OH
light curve has also been measured by \citet{silva1993}, who derived a
periodicity of 7.2 years. These data are poorly sampled, however, and
evidence for this period is not compelling.

Ground based optical observations of R22 reveal that the source is
associated with a bipolar reflection nebula \citep{allen1980}. The
irregular lobes of the nebula, each of size 2\farcs5 are separated
by 3\farcs5 at position angle 30\fdg~Nearly identical spectra
identify them as reflection nebula. Analysis of the reflection spectra
in H$\alpha$ by \citet{allen1980} shows that R22 is unusual among OH/IR stars: the nebula is
illuminated by a hidden A2 Ie star, not a more typical late M-type star. 
The spectral type is based on the presence
of Fe II absorption, the absence of He I and the weakness of the
Balmer absorption lines.

The optical morphology is strikingly confirmed in wideband continuum
Hubble Space Telescope (HST) observations centered at 606 nm, and in
associated imaging in H$\alpha$, and [SII] \citep{sahai1999}.  These high
resolution optical images show that Roberts~22
consists of two bright ``butterfly wing'' lobes of size 2\farcs5
with intersecting loops of 2\farcs5, separated by a dark equatorial
band hiding the central star. The nebular morphology is characterized
by a high degree of point symmetry, with the exception of a ``northern spur'' -- a region of decreased emission extending from the waist into the northern lobe. \citet{sahai1999} propose that the
symmetry may result from the interaction of collimated bipolar flows
with an asymptotic giant branch envelope. The authors find that the
radial emission profile is not consistent with a r$^{-3}$ profile
expected for time-invariant mass loss. 

R22 was initially detected in the near-infrared by \citet{frogel1975}.
\citet{allen1980} confirmed the near-IR properties, and explain the
infrared observations by a hot stellar continuum, reddened by
8--10$^m$. By applying this reddening to star of spectral type A2 Ib,
they obtained a distance of 1.8 kpc (although they adopted 2 kpc for
calculations) and a luminosity of $2\times10^4$ L$_\sun$. We adopt the
distance of 2 kpc in this paper, thus establishing an angular scale of
0.01 pc (2000 AU) per arc second.

\citet{allen1980} calculated a mass loss velocity of 80 \kms, from the P Cygni H$\alpha$ profile, and
suggest that the line of sight to R22 is 20-30\arcdeg \ below the
equatorial plane. $^{12}$CO (J=2-1) was detected by
\citet{bujarrabal1991} with a central velocity of -0.1 \kms~and a
velocity of expansion of $\sim35$ \kms. This CO emission is clearly
associated with R22, rather than background emission along the
galactic plane.

Initial interferometric observations of the 1612 MHz line at the
Australia Telescope Compact Array (ATCA) indicated an unresolved
source (2-3\arcsec \ diameter) with a clear velocity gradient
suggesting rotation (Goss, 1995, unpublished observations).
Subsequent ATCA observations at 1665 MHz and 1667 MHz are reported by
\citet{sahai1999} and by \citet{caswell1998} in a survey of 200 OH
masers. The overall kinematic model proposed by \citet{sahai1999} is
that of radially expanding gas at low latitudes with V$_{exp}$~$\sim$~30 \kms \ for OH and a systemic velocity of 0 to -6 \kms. They inferred a mass loss rate of 7 $\times$ 10$^{-5}$ M$_\sun$ year$^{-1}$ (from CO observations) to 10$^{-4}$ M$_\sun$ year$^{-1}$ (from dust observations),
 and estimate the mass of dust in cavity walls to be $\sim$0.3 M$_\sun$. They
 find a timescale for expansion $<$ 500 years. The authors also
proposed that the southern lobe is closer to the observer, as the
northern lobe is $\sim$0.5$^m$ fainter.

The exact kinematics of the circumstellar
material of R22 and the evolutionary status of the central star remain
important subjects of interest. In this paper, we report on
high-resolution imaging observations of the OH maser emission towards
R22 using data taken with the ATCA at 1612 MHz, 1665 MHz and 1667 MHz.
These observations reveal the morphology of the OH emission at high
resolution and allow an investigation of the maser kinematics. The
kinematic properties of the source support its classification as a
proto-planetary nebula (PPN) \citep{sahai1999}. An astrometric
solution for R22 has been obtained, which registers the OH maser
emission with respect to the high-resolution HST images published by
{\citet{sahai1999}.

We describe new radio observations in Section
\ref{sec:observations} and discuss the results of these observations
in Section \ref{sec:results}. We present an astrometric solution in Section \ref{sec:astrometry} and discuss the resulting conclusions in Section \ref{sec:star}. The discussion and conclusions are given
in Section \ref{sec:discussion} and Section \ref{sec:conclusions}
respectively.

\section{Observations}
\label{sec:observations}

Observations of R22 were made using the ATCA, located at Narrabri,
NSW, Australia on February 1 and 4, 1997, as summarized in
Table~\ref{observations}. Two 4 MHz bands were observed, centered at
1612.00 MHz (containing the OH line at rest frequency 1612.231 MHz)
and 1666.00 MHz (containing the OH main lines at rest frequencies
1665.402 MHz and 1667.359 MHz). The effective velocity resolution of
the observations was 0.44 \kms. The total observing time was 28 hours,
with the major portion of the time (15 hours) allocated to
observations at 1612 MHz. The ATCA was in configuration 6A, with a
maximum antenna spacing of approximately 6 km. For unit robust
weighting \citep{briggs1995}, full tracks at 1612 MHz provided a
resolution of $7\farcs5 \times 6\farcs2$ (PA = $-58\deg$) while the
shorter 1666 observation resulted in a resolution of $20\farcs4 \times
5\farcs1$ (PA = $-13.4\deg$).


The data were edited, calibrated and imaged using the Astronomical
Image Processing System (AIPS) maintained by the National Radio Astronomy Observatory (NRAO). The
observations at 1612 MHz were subject to severe interference from a
Glonas satellite broadcasting at 1608 MHz. Careful editing of the data
was required to eliminate these effects. The 1612 MHz and 1666 MHz
data sets were self-calibrated in phase on the channels with strong
maser emission. These solutions were applied to all remaining
channels. Self-calibration changed the positions of our sources by
$\lesssim$ 0\farcs1. The images were deconvolved using the Clark
CLEAN algorithm down to a threshold level of five times the rms
noise. The AIPS task JMFIT was used to fit the centroid of emission in
each velocity channel in the image plane. For typical line features
with a peak flux density of 10 Jy beam$^{-1}$, the relative error in
the determination of positions is $\sim$20 mas, assuming the error
model of \citet{kogan1997}. The position error contributed by the
phase calibrator J0823-500 is limited by an uncertainty in its VLBI
position of 2-3 milliarcseconds (mas) (J. Reynolds, private
communication).

In the strongest channels, the images are dynamic-range limited at the
level of approximately 1000:1 at 1612 MHz and 1665 MHz, and 100:1 at
1667 MHz.  The observational parameters are summarized in Table
\ref{results}; the quoted rms noise values (20 mJy beam$^{-1}$ at 1612
MHz and 28 mJy beam$^{-1}$ at 1665 MHz) were measured in channels
without strong emission.

\section{Results}
\label{sec:results}

\subsection{Continuum Results}

A continuum image was constructed from the line free channels in the
1612 MHz data. The image shows two weak continuum sources
$\sim$10\arcmin \ to the west of the position of R22, with integrated
flux densities of 0.5 Jy and 0.2 Jy. The sources, located near
$\alpha= $10$^h$ 20$^m$ 15$^s$, $\delta$=~-58\arcdeg~04\arcmin \
(J2000), are part of the \ion{H}{2} region G 284.6-0.5, imaged at 6 cm by
\citet{shaver1970}. No continuum emission was detected at the position
of R22 to within a 3$\sigma$ limit of 7 mJy beam$^{-1}$.

\subsection{Line Results}
\label{sec:lineresults}


At 1612 MHz (see Figure~\ref{lines}a) the integrated line emission
profile of R22 shows a multiple-peaked spectrum with a peak flux
density of 42 Jy beam$^{-1}$ and a velocity range of 50 \kms, in
agreement with the earlier observations of \citet{allen1980}.  The
integrated line emission profile at 1665 MHz, shown in
Figure~\ref{lines}b, is clearly double-peaked, but asymmetric, with
the blue-shifted emission predominating, in agreement with the 1612 MHz emission. 
The 1665 MHz emission has a
peak flux density of 50 Jy beam$^{-1}$ and a velocity range similar to
the 1612 MHz line. The weaker 1667 MHz integrated line emission, which
has a maximum of 2.9 Jy beam$^{-1}$ is shown in Figure \ref{67line}. 

Recent measurements by \citet{caswell1998} found 1665 MHz and 1667 MHz
emission at $\alpha$= 10$^h$ \ 21$^m$ \ 33\fs87 $\delta$=
-58\arcdeg~05\arcmin~47\farcs6 ($\pm$ 0\farcs4, J2000). This
determination agrees with the position of the peak flux density at
1665 MHz and 1667 MHz from the observations reported here, as
summarized in Table \ref{results}. Caswell measured a peak flux
density of 2.1 Jy for the peak 1667 MHz emission, in good agreement
with our observed flux density of 2.9 Jy beam$^{-1}$. The peak flux
density of the 1665 MHz line, as reported by \citet{caswell1998}
(about 70 Jy beam$^{-1}$), is quite uncertain due to the large
correction for primary beam attenuation.

\placetable{results} 

At 1612 MHz the locations of the centroids of the maser emission in
each channel describe an arc, as shown in Figure \ref{centroids}a. This
arc covers a position angle (PA) from 310\arcdeg \ to 160\fdg~At
the adopted distance of 2 kpc \citep{allen1980,sahai1999},
the maser emission spans approximately 2800 AU. The velocity increases
as PA decreases around the arc, from -35 \kms \ in the NW to -5 \kms \
in the SE. At 160\arcdeg \ the velocity--PA correlation changes sign,
with velocity increasing from 0 \kms \ at 170\arcdeg \ to 15 \kms \ at
210\arcdeg. The pattern for the 1665 MHz emission (shown in Figure
\ref{centroids}b) is similar to 1612 MHz, with the same velocity--PA
change of sign at 170\arcdeg.  These relations are plotted in Figure \ref{thetav}. The brightest 1667 MHz maser emission
is located near the center of the 1665 and 1612 MHz arcs, but does not
show any clear velocity-position angle dependence.



\subsection{Optical and Radio Morphology}
\label{sec:morphology}


\subsubsection{Astrometry}
\label{sec:astrometry}

We aligned the HST image of R22, as obtained from the HST archive, to
the radio reference frame using guide stars from the USNO-A2.0
catalog. We used mosaiced HST observations of R22, with the object
located on the Planetary Camera chip and containing 32 stars from the
UNSO-A2.0 catalog, all fainter than 12 magnitude. Table \ref{astrometry} lists the stars used, along with residuals and blue and red magnitudes. Figure \ref{astro.image} shows the stars plotted over the HST image.  The USNO-A2.0
catalog has absolute astrometric calibration based on the
International Celestial Reference Frame, through the Hipparcos Tycho
catalog, and is in agreement with the radio reference frame to within
0\farcs26 \citep{deutsch1999}. \citet{deutsch1999} found that with
sufficient reference stars, the position of current (epoch 2000)
optical sources not found in USNO-A2.0 can be achieved to a
(1$\sigma$) accuracy of 0\farcs35. The dominant source of error is
small-scale proper motions of reference stars since the 1950-era
observations.  We investigated several sources of additional error
which may have affected our solution, including the proper motion of
nearby stars and the uncertainty in relative chip positions. As
recommended by Deutsch (1999) we excluded confused stars (such as
stars resolved into optical binaries by HST) and stars with obvious
proper motions from our fit. We compared solutions for stars on a
single chip to solutions over all four chips (including a solution primarily using stars on the WPC) and found no discrepancy
in star positions across the chip boundaries.   Therefore the alignment
of the HST and ATCA observations of R22, shown in Figure \ref{cont1}
and \ref{cont67}, has an accuracy of $\pm$ 0\farcs35 (1$\sigma$).

The 1612 and 1665 MHz emission features are associated with the
northern lobe as shown in Figure \ref{cont1}. The 1612 MHz emission
describes a circular arc coincident with regions of fainter optical
emission in the waist of the bipolar outflow and in the ``northern
spur'', clearly visible as a region of decreased intensity in Figure \ref{greys}. The 1665 MHz emission (see Figure \ref{cont1}b) is located
along the same arc. The 1667 MHz emission is less precisely located,
due to lower signal to noise but is clearly not confined to the same
region (see Figure \ref{cont67}). We do not observe the equatorial spread of 1667 MHz emission
shown in Figure 3b of \citet{sahai1999}; rather the emission seems
approximately bounded by the extended emission of the nebulae, with
the brightest masers (2.8 Jy) located near $\alpha$= 10$^h$ 21$^m$
33\fs9, $\delta$= -58\arcdeg~05\arcmin~48\arcsec~(J2000) (see Figure
\ref{cont67}).

The alignment of the HST and OH maser observations as derived here,
differs from that adopted by \citet{sahai1999}. In the absence of an
astrometric solution to register the OH and optical emission,
\citet{sahai1999} assumed that the maser emission was aligned with the
geometric center of the nebula, confined to the equatorial waist
(Sahai, private communication). Their spatial alignment is therefore
uncertain at, or above, the arcsecond level.




\subsubsection{The position of the star}
\label{sec:star} 

Without full spatial sampling, the arc of emission at 1612 MHz and
1665 MHz, shown in Figure \ref{centroids} can be fit by a circle with radius 0\farcs7
centered at $\alpha$= 10$^h$ 21$^m$ 33\fs95 
 $\delta$= -58\arcdeg~05\arcmin ~47\farcs3 
We do not believe this position is the most likely location of the star. As stated by \citet{bowers1991} and \citet{hekkert1992} the most likely position for an obscured central star is the midpoint between the red- and blue-shifted maser emission. However this method is only reliable if the maser distribution is fully sampled in velocity, and is approximately symmetric in appearance. For the case of R22, the distribution on the sky is clearly non-symmetric (see
Figure \ref{centroids}), and it is quite likely that maser emission has not been
detected from the full range of velocities. Given the point symmetry in the nebulae structure, a more likely estimate is the center of the optical nebula. In the Hipparcos reference frame this suggested stellar position is $\alpha$=
10$^h$~21$^m$~33\fs85, $\delta$= -58\arcdeg~05\arcmin~47\farcs8 
 (J2000). It should be noted that while we have accurately determined the optical-radio alignment, choosing a center of a non-symmetric nebula is a matter of opinion. The location marked by a cross in Figure \ref{centroids} and diamonds in Figures \ref{cont1}-\ref{cont67} is the same location (relative to the nebula) proposed by \citet{sahai1999}, but aligning the HST image to the Hipparcos reference frame has shifted the optical coordinates by $\Delta
\alpha$= -1\farcs3, $\Delta \delta= -0\farcs7$.

\section{Discussion}
\label{sec:discussion}

The radio-interferometric images reported here are the first at 1612
MHz, and demonstrate a clear arc-like morphology in both the 1612 MHz
and 1665 MHz OH maser emission. The observations have sufficient
sensitivity to permit accurate measurements of the position of the
centroid of the emission and the variation of maser component velocity
across the spectrum. These signatures can be used to explore the
kinematics of the circumstellar material in the region in which the OH
masers are located in R22.

Double-peaked total-power spectra, such as those observed for R22
(Figs. 1-2), are commonly associated with spherical mass outflows in a
stable maser emission region \citep{bowers1991}, modulo deviations
explained by localized mass loss and density variations
\citep{silva1993}. However, such outflows produce a characteristic
radius-velocity signature, described by
$\theta(V_{LSR})=\theta_R \sqrt{1-{{(V_{LSR}-V_*)}^2 \over{V_e^2}}}$, where
$V_{LSR}$ is the local standard of rest (LSR), $\theta_R$ is the radius of the shell, $V_*$ is the velocity of the central star, and $V_e$ is the expansion velocity. We plot the
measured radius-velocity relationship for R22 as derived from the 1612
MHz and 1665 MHz centroid position in Figure 7, assuming the stellar
position at the optical center of the nebula, as described in the
preceding section. A spherically symmetric outflow is clearly not
consistent with these data.

As discussed above, there is also a systematic variation of velocity
with position angle along the arcs of emission at both 1612 MHz and
1665 MHz. This also argues strongly against a spherically symmetric
outflow. The kinematic signatures observed for R22 could be produced
by a variety of outflow models, perhaps in combination with local
masing conditions which vary across the source. 

Kinematic modeling of the maser emission was attempted, by
fitting to parametrized ellipsoidal models, similar to those described
by \citet{bowers1991}. Such fitting is sensitive to the assumed stellar
position, the degree of completeness of the velocity and spatial
sampling, and the assumptions regarding maser transport. The models
employed assumed a regular kinematic flow with one component. Maser
intensity was derived directly from the velocity coherence along the
line of sight, and no allowance was made for local masing conditions,
density or turbulence. No conclusive model has been obtained, and
further work is planned in this area.

We can conclude from our study of the kinematic properties of the source that R22 does not show characteristics of a classical OH/IR star. The highly asymmetric OH maser
emission supports the classification as a PPN, as clearly supported by the
HST observations \citep{sahai1999}.

The accurate relative astrometry reported here shows a striking
spatial association between the arcs of OH maser emission at 1612 MHz
and 1665 MHz and the bipolar optical emission, especially in the
region of the ``northern spur'', shown in Figure \ref{cont1}. 

\section{Conclusions}
\label{sec:conclusions}

We have presented radio-interferometric images of the OH maser
emission towards R22, obtained using the ATCA, in the transitions at
1612 MHz, 1665 MHz and 1667 MHz. The principal conclusions of this
work are:

\begin{itemize}

\item{The OH maser emission at 1612 MHz and 1665 MHz has an arc-like
morphology, with a clear, systematic dependence of maser position and
velocity across the spectrum. The kinematic signatures are not
consistent with a classical OH/IR star with a spherically symmetric
outflow.}

\item{The OH maser emission and the HST optical image have been
astrometrically aligned to within $\pm$ 0.35\arcsec \
(1$\sigma$). This defines the exact location of the OH maser emission
in the equatorial waist of the optical bipolar outflow.}

\item{The maser kinematics support the classification of this object
as a PPN as suggested by \citet{sahai1999}.}

\end{itemize}

\acknowledgments

The Australia Telescope is funded by the Commonwealth of Australia for
operation as a National Facility managed by CSIRO. The National Radio
Astronomy Observatory is a facility of the National Science Foundation
operated under a cooperative agreement by Associated Universities,
Inc. This research has made use of the NASA Astrophysics Data System
Abstract Service.

 We would like to thank A. Silva for suggesting new observations of
 Roberts~22, P. Diamond and M. Claussen for help in the initial stages
 of the project, J. Chapman for comments and F. Owen, L. Van Zee and
 E. Deutsch for assistance with the astrometry. We are
 especially grateful to P. Bowers for bringing his extensive
 experience to bear on R22.

\clearpage

\onecolumn

\begin{figure}
\plottwo{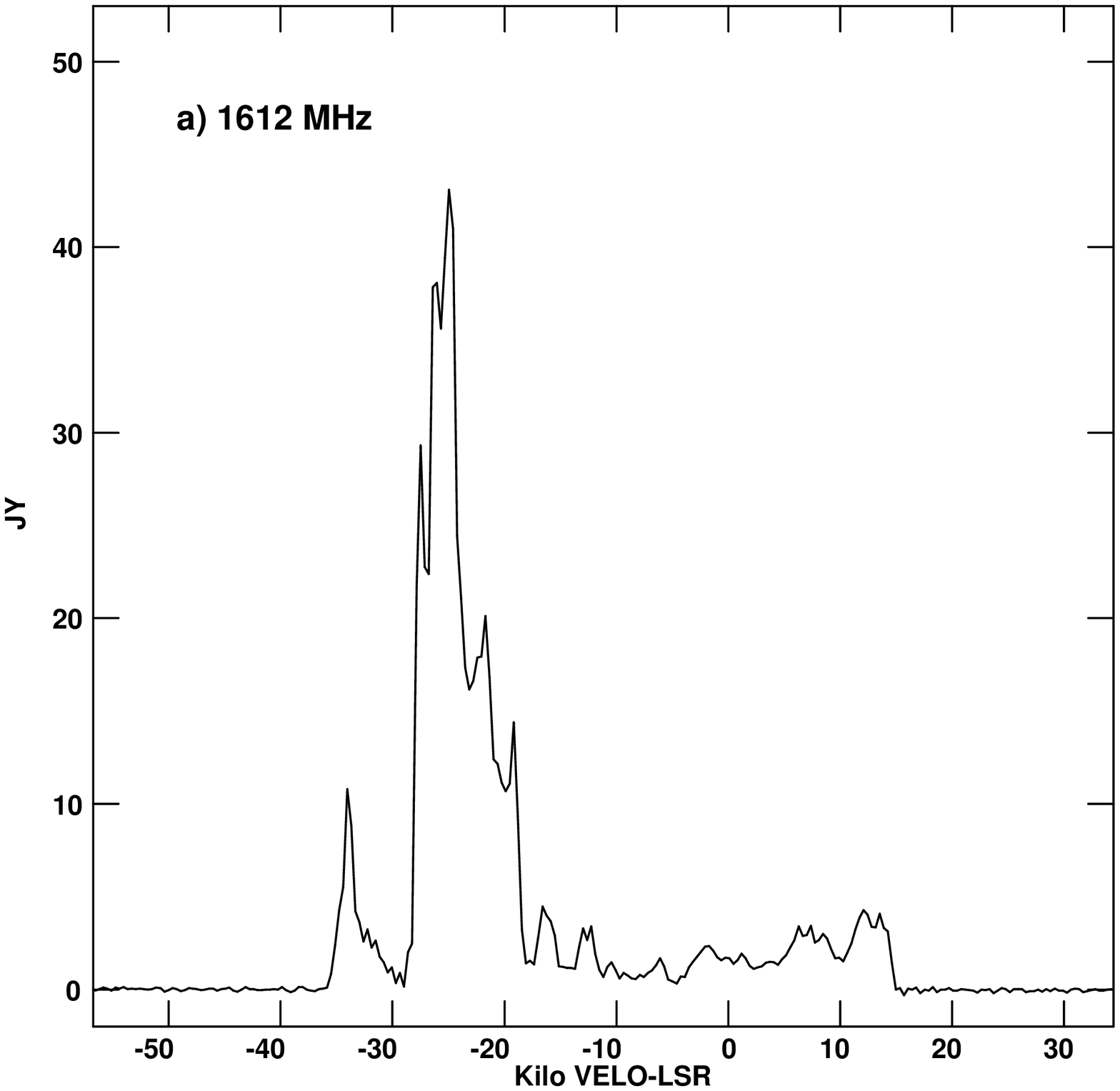}{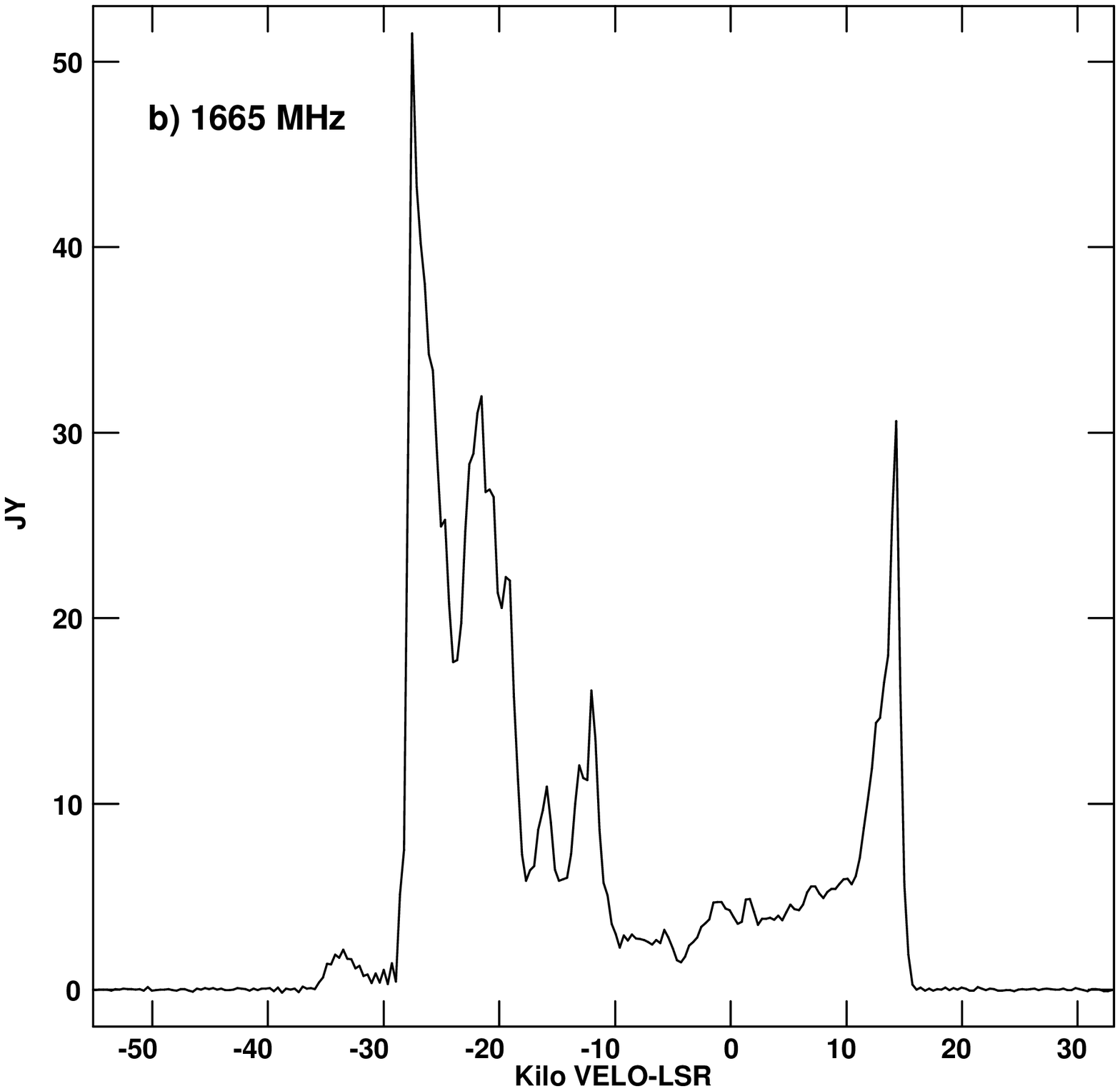}
\caption{The line emission profile of OH transitions at 1612 MHz (\ref{lines}a), resolution 7\farcs5 $\times$ 6\farcs2 PA = $-58\deg$ and 1665 MHz (\ref{lines}b) resolution 20\farcs4 $\times$ 5\farcs1 PA = $-13.4\deg$, observed with the Austrian Telescope Compact Array in February 1997. The velocity is measured with respect to the LSR. These profiles are integrated over the source.\label{lines}}
\end{figure}
\clearpage 

\begin{figure}
\plotone{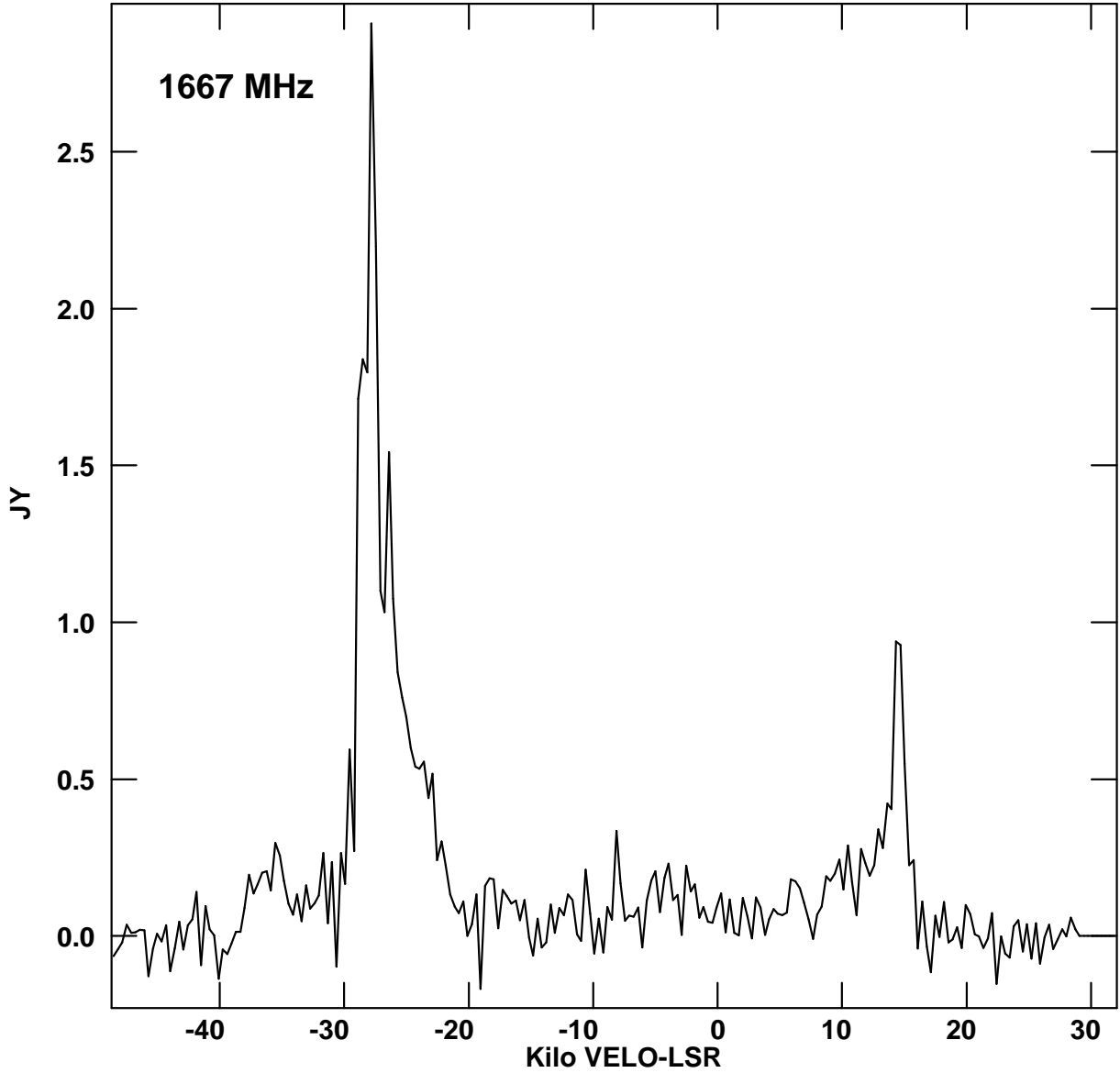}
\caption{Line emission profile of OH 1667 MHz, resolution 20\farcs4 $\times$ 5\farcs1 PA = $-13.4\deg$, observed with the ATCA February 1997. See Figure \ref{lines}. \label{67line}}
\end{figure}
\clearpage 

\begin{figure} 
\plottwo{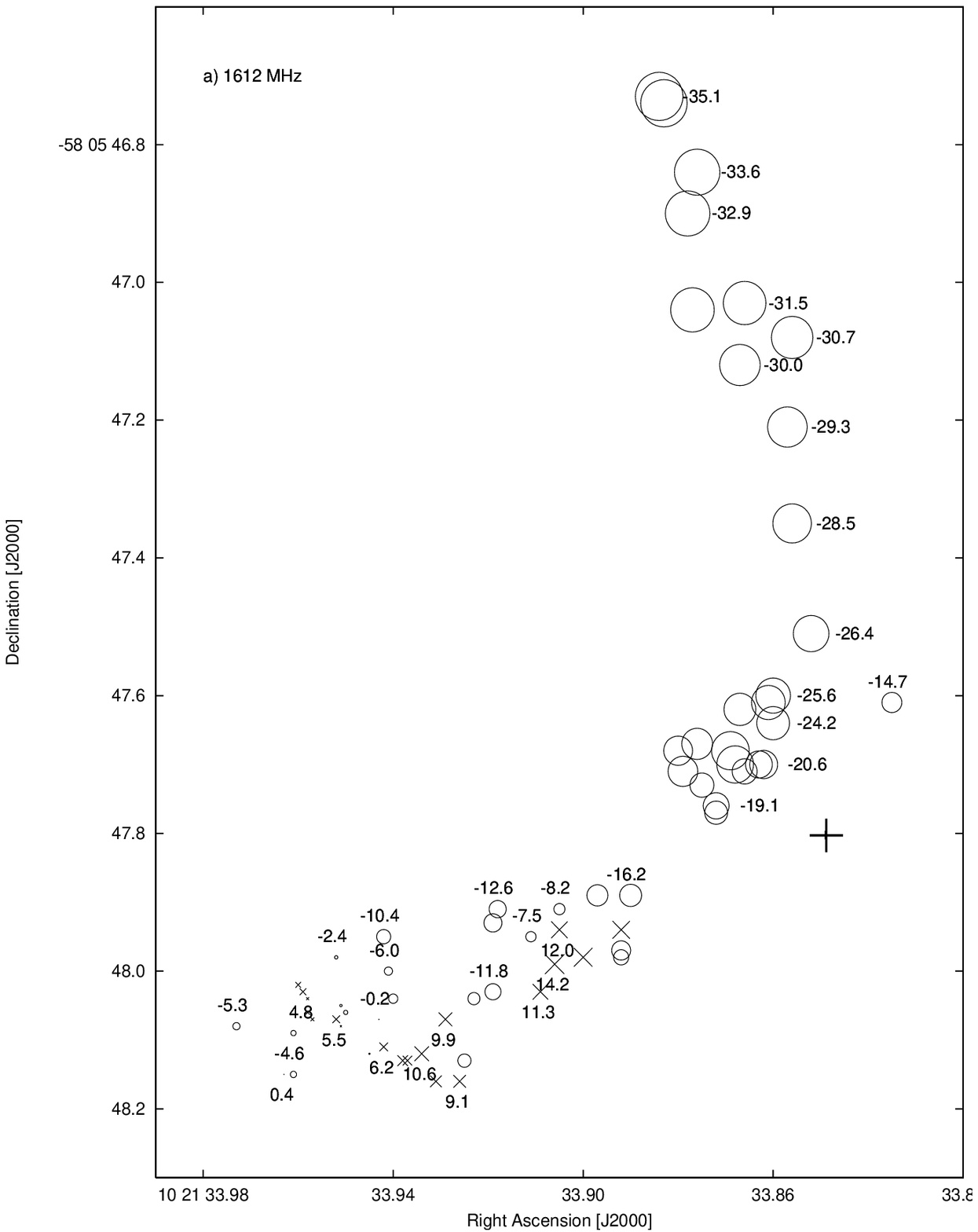}{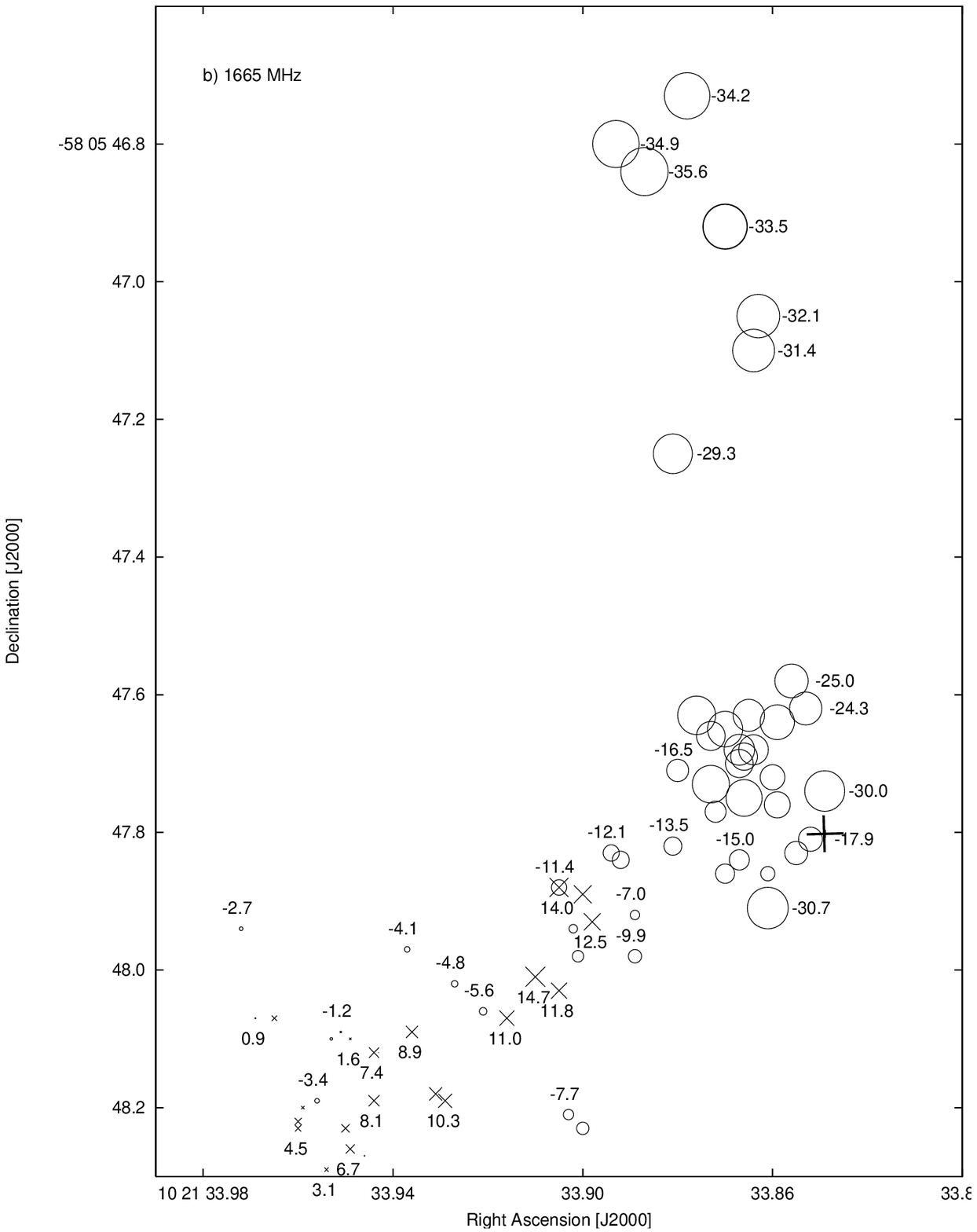}
\caption{a) 1612 MHz and b) 1665 MHz velocity centroids. The size of the points are scaled by the velocity (with respect to the LSR). Negative velocities are represented as circles (label above or to the right), positive velocities are X's (label below). The cross at $\alpha$= 10$^h$~21$^m$~33\fs85, $\delta$= -58\arcdeg~05\arcmin~47\farcs8~(J2000) marks the geometric center of the optical nebula as discussed in \S \ref{sec:star}.\label{centroids}}
\end{figure}
\clearpage

\begin{figure} 
\plottwo{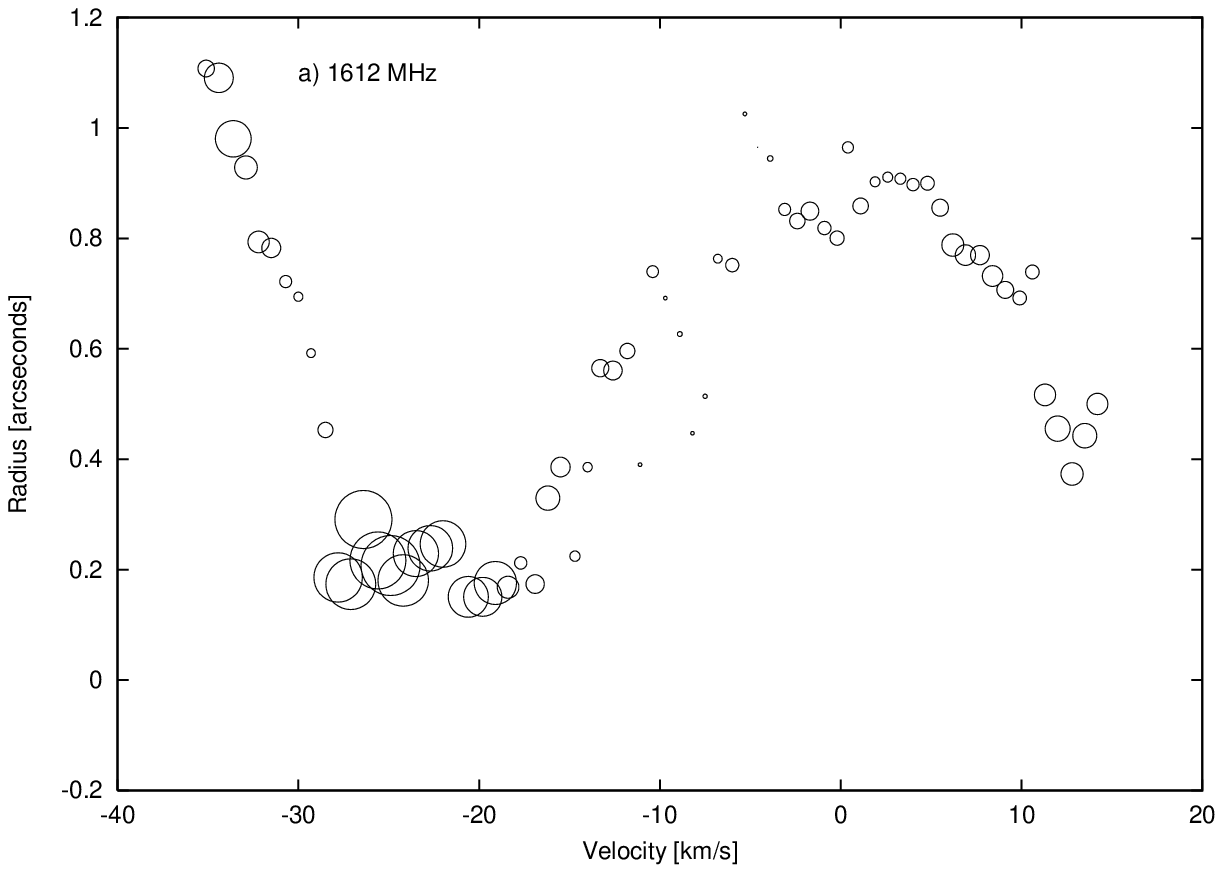}{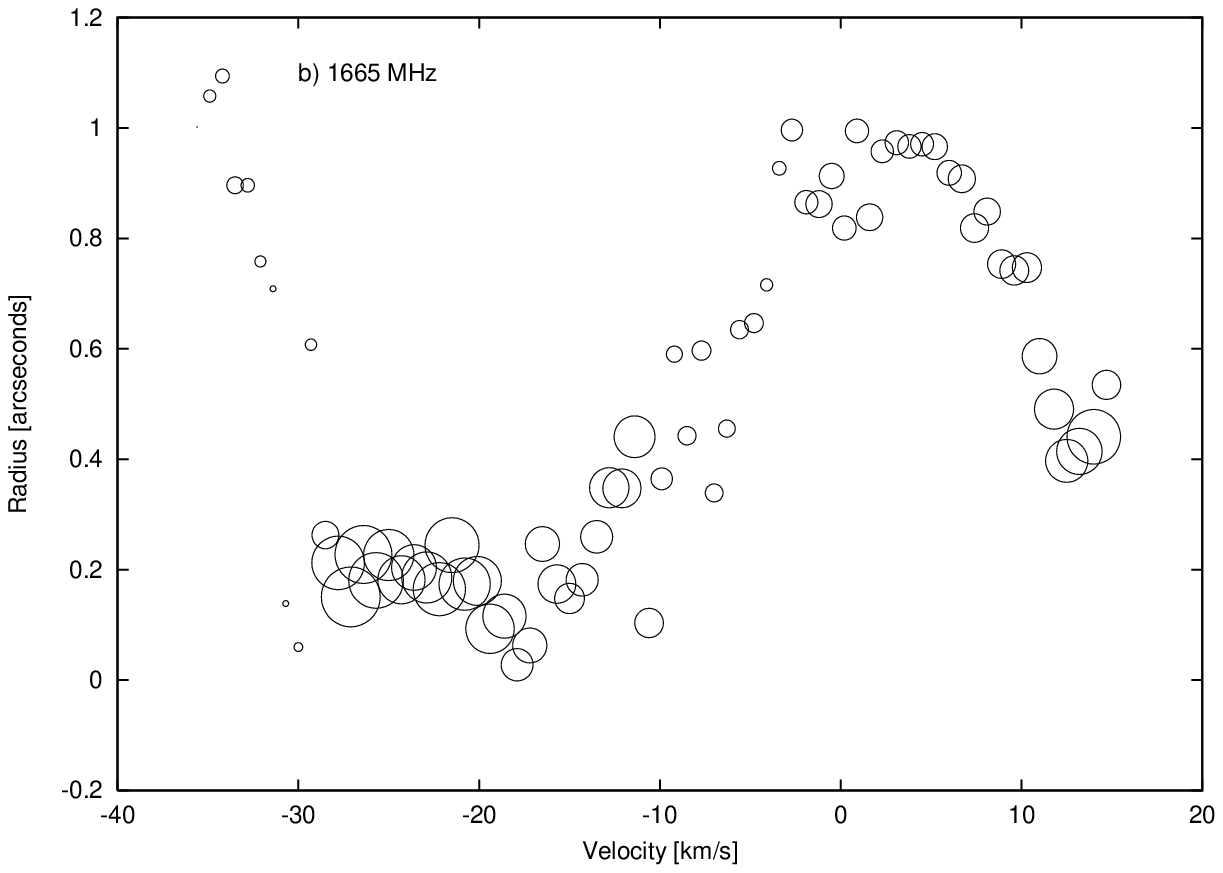}
\caption{Radius-velocity relationship for a) 1612 MHz emission and b) 1665 MHz emission with radius measured from the geometric center of the optical nebula (see Section \ref{sec:star}). The size of the symbols varies on a logarithmic scale with flux density. The largest symbol indicates a flux of 50 Jy, the 1665 MHz peak. Velocities are measured with respect to the LSR.   \label{thetav}}
\end{figure}

\begin{figure} 
\plotone{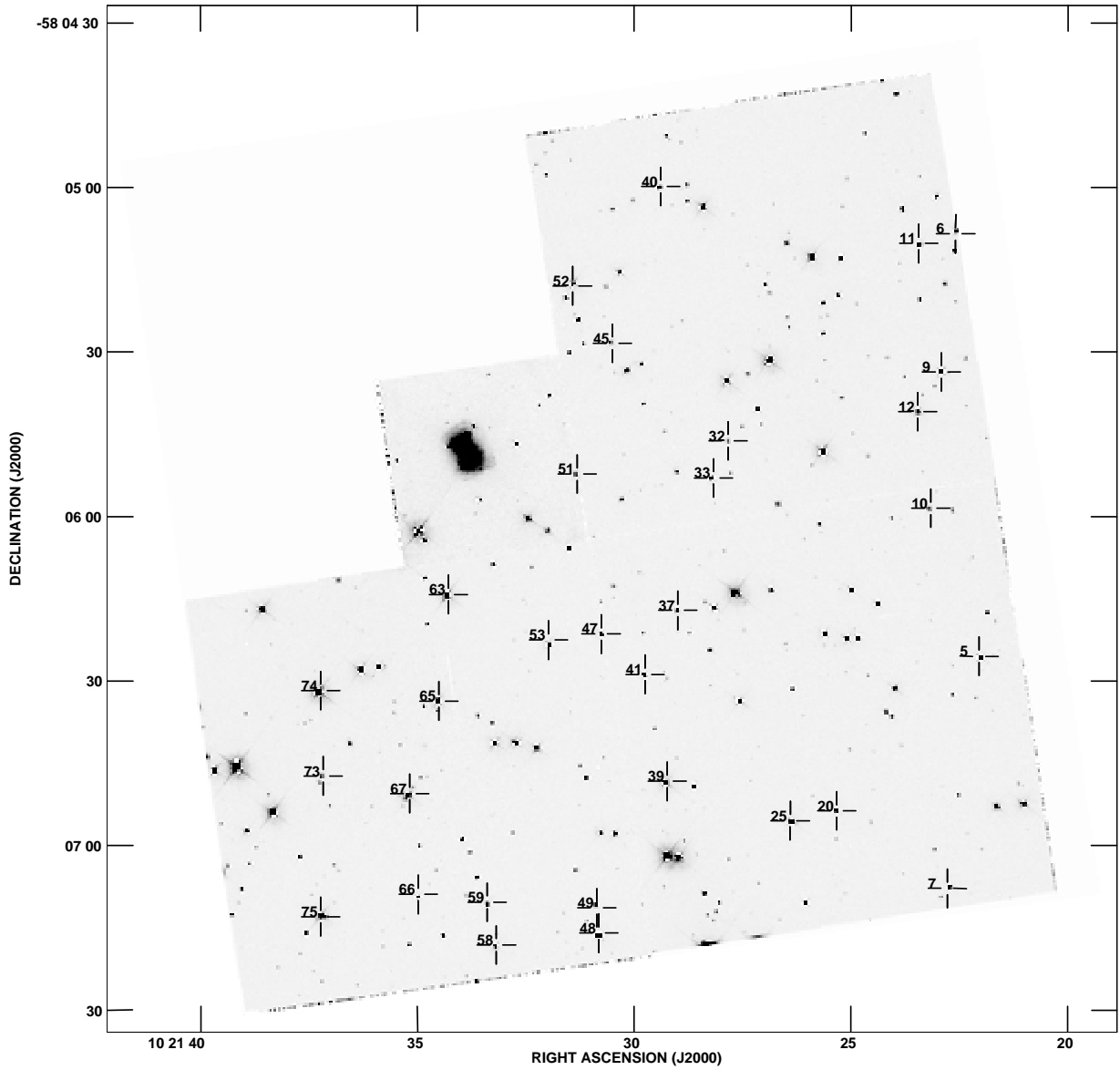}
\caption{The USNO stars used for the astrometric solution. Roberts~22 is centered on the WPC. Numbers refer to the star designation in Table \ref{astrometry}  \label{astro.image}}
\end{figure}

\begin{figure} 
\plottwo{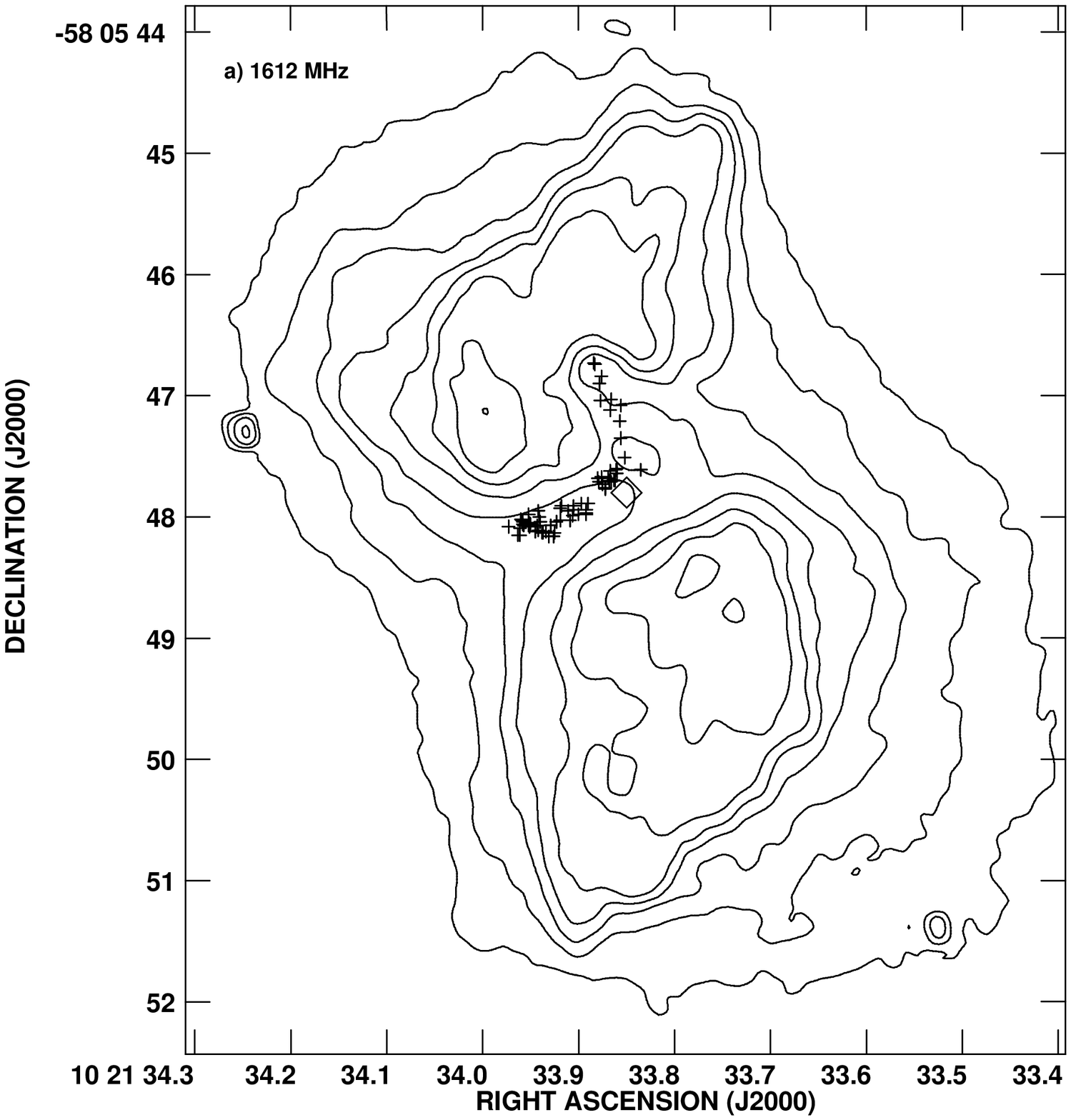}{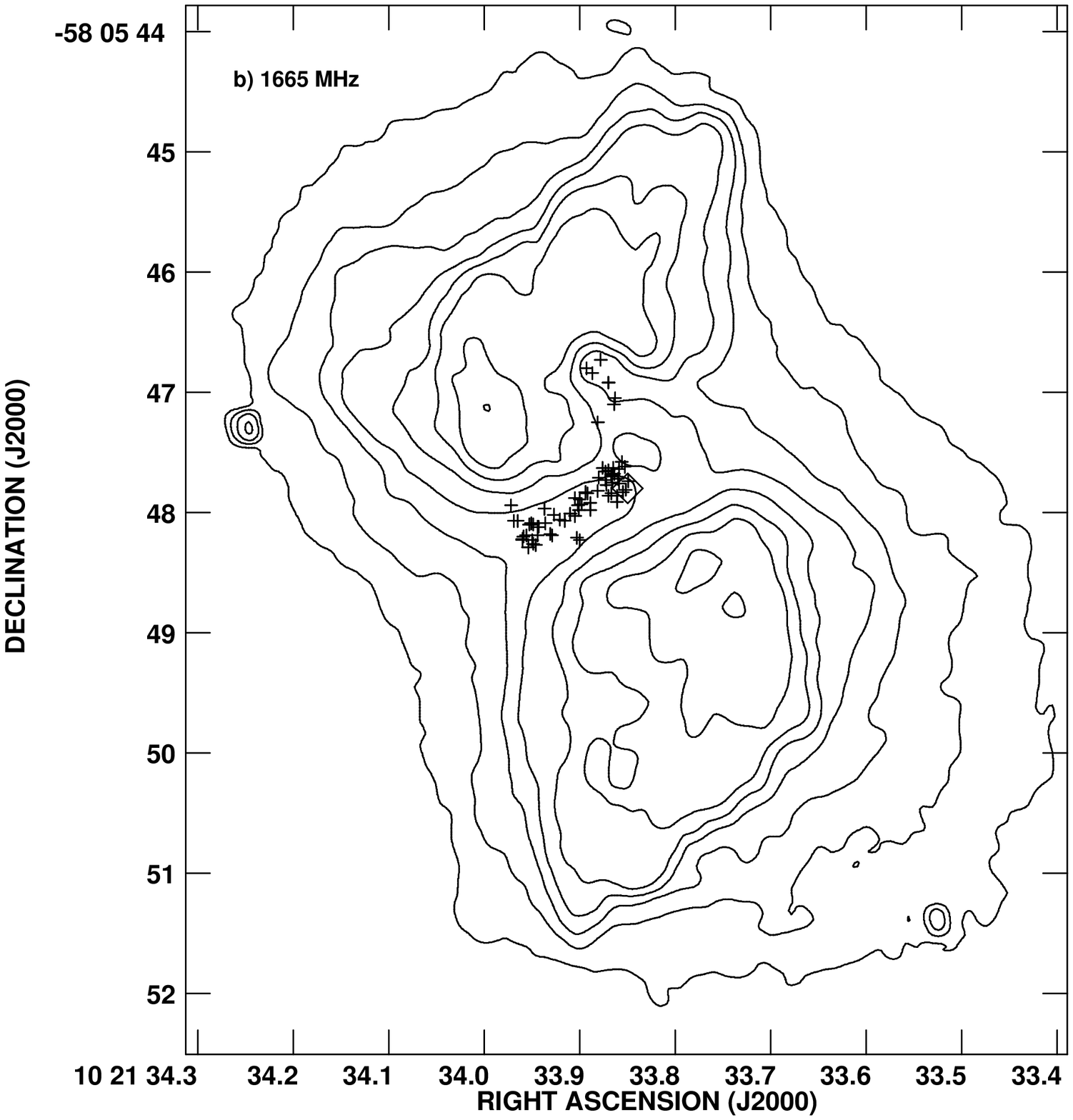}
\caption{HST H$\alpha$, observation by \citet{sahai1999} with crosses representing the a) 1612 MHz maser positions and b) 1665 MHz maser positions. The logarithmic scale contours represent 8, 16, 32, 64, 128, 256, 512 \& 1014 times the background. The diamond at $\alpha$= 10$^h$ 21$^m$ 33\fs85, $\delta$= -58\arcdeg~05\arcmin~47\farcs8~(J2000) marks the center of the optical nebula, as discussed in \S \ref{sec:star}.
\label{cont1}}
\end{figure}
\clearpage

\begin{figure} 
\plotone{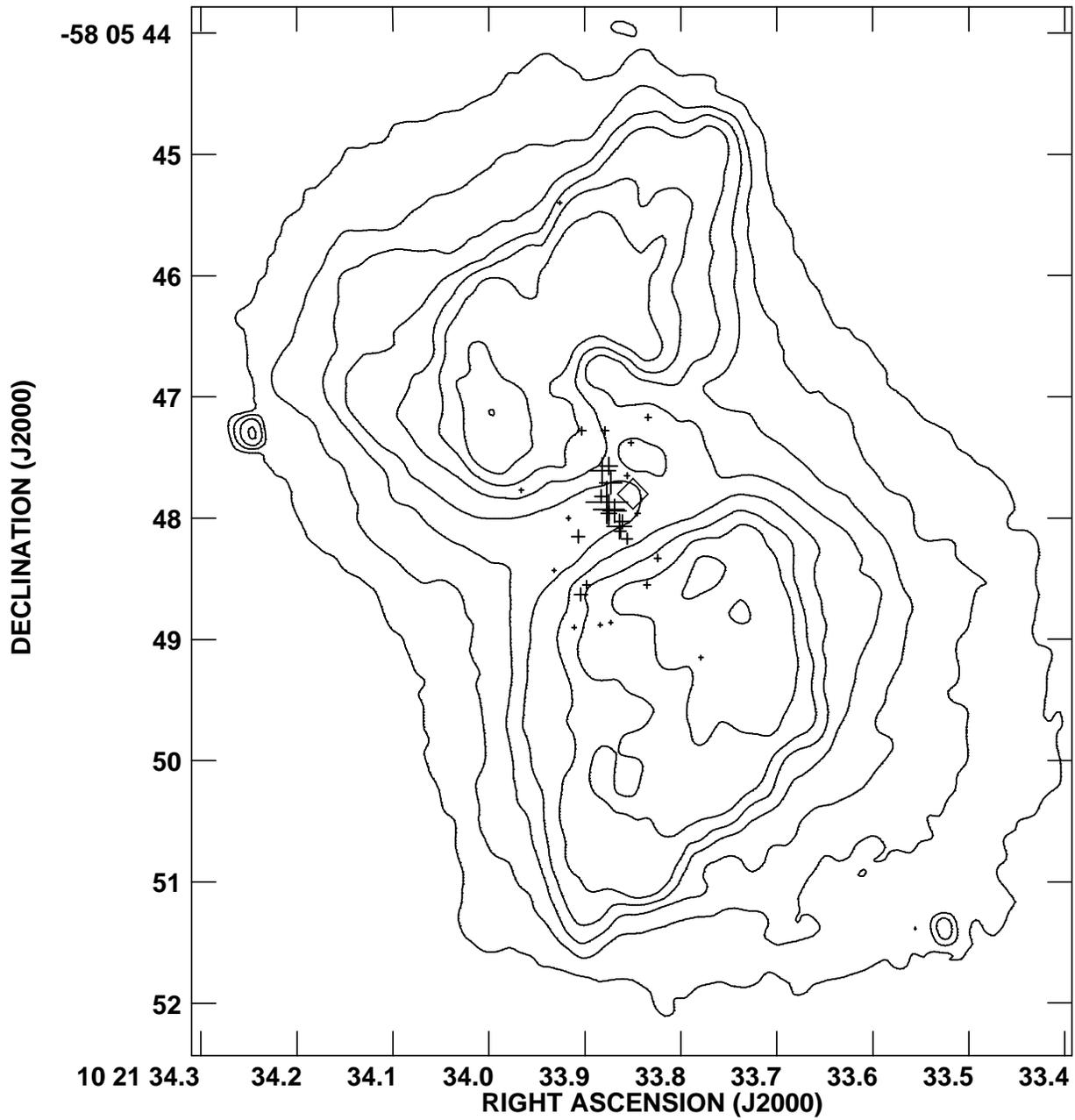}
\caption{HST H$\alpha$, observation by \citet{sahai1999} with 1667 MHz maser positions in this plot scaled by intensity. The brightest maser emission is 2.9 Jy. Plot details are discussed in Figure \ref{cont1}.
\ref{sec:star}.\label{cont67}}
\end{figure}
\clearpage

\begin{figure} 
\plotone{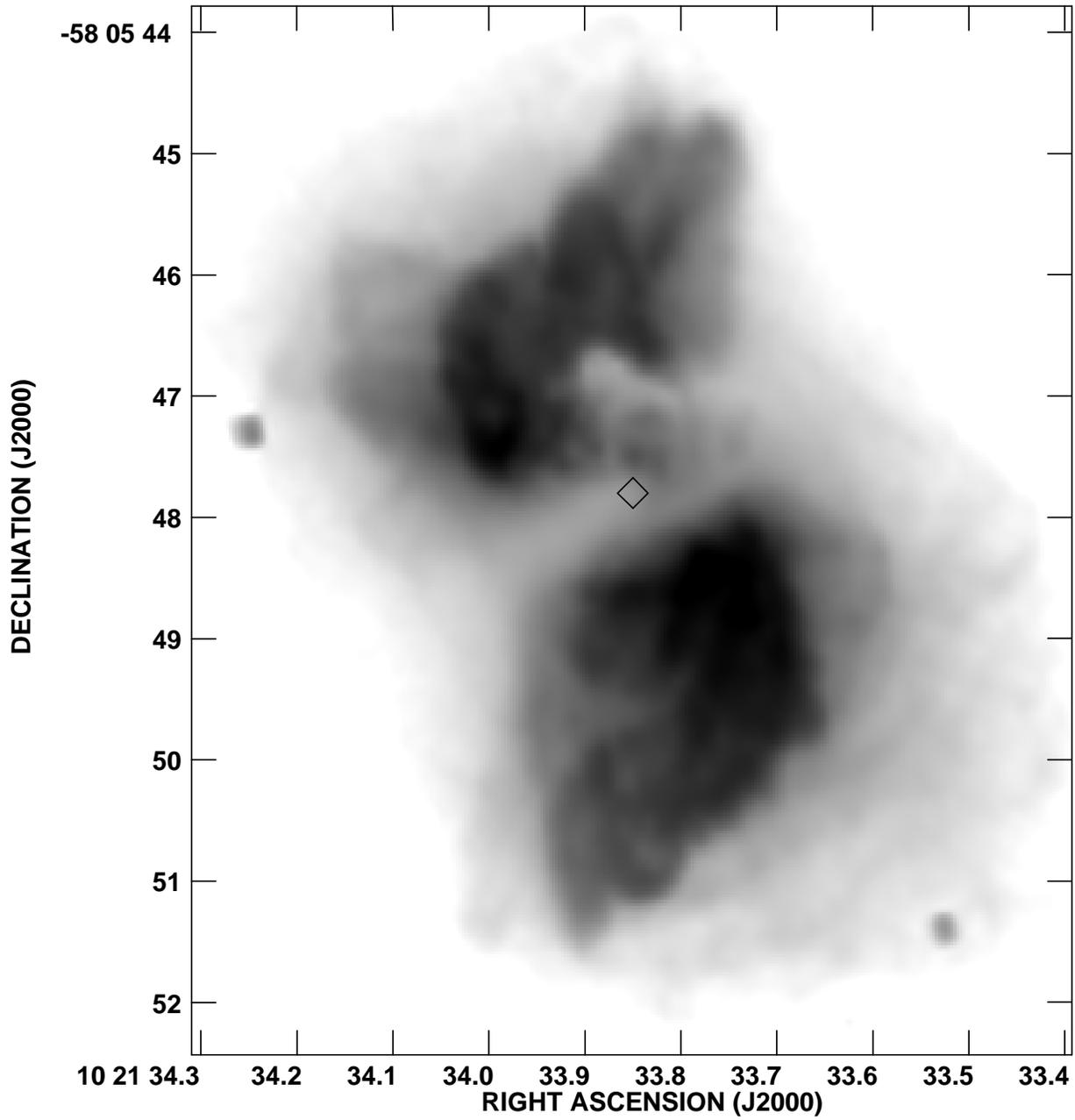}
\caption{HST H$\alpha$, observation by \citet{sahai1999}, on a logarithmic scale with the geometric center marked with a diamond as discussed in \S \ref{sec:star}. Note the ``northern spur'', an area of reduced emission extending from the waist into the northern lobe. \label{greys}}
\end{figure}
\clearpage

\begin{deluxetable}{rl}
\footnotesize
\tablecaption{ATCA OH Observations of Roberts~22\label{observations}}
\tablewidth{0pt}
\tablehead{}
\startdata
Observation Date & February 1,4 1997\\
Configuration& 6A \\
$\alpha,~\delta$ (J2000) & 10$^h$ 21$^m$ 33\fs98 -58\arcdeg~05\arcmin~38\arcsec \\
$\ell,~b$&284\fdg18  -0\fdg79\\
Total Bandwidth& 4 MHz/740 \kms \\
Number of Channels& 2048\\
Channel Separation&  1.95 kHz/0.36 km s$^{-1}$ \\
Spectral Resolution& 2.36 kHz/0.44 km s$^{-1}$ \\
Flux Density Calibrator& 1934-638, 14.5 Jy at 1612 MHz\\
Phase Calibrator& J0823-500, 7.7 Jy at 1612 MHz \\
\enddata
\end{deluxetable}

\begin{deluxetable}{rccc}
\footnotesize
\tablecaption{ATCA OH Observations of Roberts~22\label{results}}
\tablewidth{0pt}
\tablehead{&\colhead{1612 MHz}&\colhead{1665 MHz}&\colhead{1667 MHz}}
\startdata
Observation Duration&	925 min&	140 min&	140 min \\
Spatial Resolution& 
7\farcs5 $\times$ 6\farcs2& 20\farcs4 $\times$ 5\farcs1 & 20\farcs4 $\times$ 5\farcs1\\
Beam Position Angle&-58\arcdeg&   13\arcdeg &  13\arcdeg \\
RMS Noise& 10 mJy beam$^{-1}$&  24 mJy beam$^{-1}$&  24 mJy beam$^{-1}$ \\
Peak Flux Density& 42.2 Jy beam$^{-1}$ & 50.1 Jy beam$^{-1}$ &  2.88 Jy beam$^{-1}$ \\
Location of Peak: $\alpha$&  10$^h~$21$^m~$33\fs86 & 10$^h~$21$^m~$33\fs87 & 10$^h$ 21$^m$ 33\fs9\\
(J2000) $\delta$& -58\arcdeg~05\arcmin~47\farcs6 & -58\arcdeg~05\arcmin~47\farcs7 &-58\arcdeg~05\arcmin~47\farcs9 \\
Velocity of Peak&-25.0 \kms & -27.5 \kms & -27.8 \kms \\

\enddata
\end{deluxetable}

\begin{deluxetable}{lrrrrccc}
\footnotesize
\tablecaption{USNP Stars \label{astrometry}}
\tablewidth{0pt}
\tablehead{\colhead{Ref.\tablenotemark{1}}&\colhead{RA: h m s~\tablenotemark{2}~}&\colhead{Residuals \arcsec~\tablenotemark{3}}&\colhead{DEC:\arcdeg~\arcmin~\arcsec}&\colhead{Residuals \arcsec}&\colhead{Red $^{m}$}&\colhead{Blue $^{m}$}}
\startdata
5 & 10 21 22.0420 & -0.10 & -58 06 25.480 	&-0.23 & 17.2 & 18.6\\
6 & 10 21 22.5880 & -0.18 & -58 05 08.470 	& 0.40 & 16.9 & 18.2\\
7 & 10 21 22.7760 & -0.58 & -58 07 07.830 	& 0.33 & 17.4 & 19.1\\
9 & 10 21 22.9127 & 0.29 & -58 05 33.650 	&-0.08 & 16.9 & 18.3\\
10 & 10 21 23.1640 &  0.75 & -58 05 58.510 	&-0.27 & 17.7 & 19.5\\
11 & 10 21 23.4380 & -0.23 & -58 05 10.220 	&-0.32 & 17.4 & 19.3\\
12 & 10 21 23.4653 &  0.05 & -58 05 40.870   	&-0.02 & 16.8 & 18.0\\
20 & 10 21 25.3260 & -0.03 & -58 06 53.700   	& 0.06 & 17.7 & 19.8\\
25 & 10 21 26.3993 & -0.07 & -58 06 55.470   	& 0.04 & 16.9 & 18.0 \\
32 & 10 21 27.8387 & 0.01 & -58 05 46.210   	&-0.07 & 17.8 & 19.9\\
33 & 10 21 28.1720 & 0.43 & -58 05 52.970 	& 0.02 & 17.6 & 19.0\\
37 & 10 21 28.9907 & 0.01 & -58 06 17.150 	& 0.18 & 17.5 & 19.3\\
39 & 10 21 29.2400 & 0.62 & -58 06 48.310 	&-0.16 & 15.6 & 17.4\\
40 & 10 21 29.3893 & -0.09 & -58 04 59.840   	&-0.28 & 17.5 & 18.9\\
41 & 10 21 29.7473 & -0.01 & -58 06 28.800   	& 0.14 & 17.7 & 19.5\\
45 & 10 21 30.4973 & 0.15  &  -58 05 28.430  	& 0.02 & 17.7 & 19.3\\
47 & 10 21 30.7627 & 0.06 & -58 06 21.450 	&-0.06 & 17.7 & 19.3\\
48 & 10 21 30.8133 & -0.17 & -58 07 15.990 	&-0.25 & 17.0 & 18.1\\
49 & 10 21 30.8613 & 0.22 & -58 07 11.400  	& 0.57 & 17.7 & 19.0\\
51 & 10 21 31.3127 & 0.26 & -58 05 52.290  	& 0.72 & 17.4 & 19.0\\
52 & 10 21 31.4233 & -0.58 & -58 05 17.940 	& 0.40 & 15.9 & 17.2\\
53 & 10 21 31.9747 & -0.22 & -58 06 22.560 	&-0.57 & 17.1 & 18.6\\
58 & 10 21 33.1773 & 0.09 & -58 07 18.150 	&-0.14 & 17.3 & 18.6\\
59 & 10 21 33.3880 & -0.31 & -58 07 10.370 	&-0.03 & 17.4 & 18.8\\
63 & 10 21 34.2860 & 0.14 & -58 06 14.230 	&-0.03 & 15.9 & 16.8\\
65 & 10 21 34.5047 & -0.08 & -58 06 33.660  	& 0.11 & 15.7 & 18.2\\
66 & 10 21 34.9773 & -0.16 & -58 07 08.930 	&-0.14 & 17.5 & 19.7\\
67 & 10 21 35.1753 & 0.05 & -58 06 50.530 	&-0.21 & 16.4 & 17.1\\
73 & 10 21 37.1733 & -0.02 & -58 06 47.310   	& 0.25 & 17.3 & 18.6\\
74 & 10 21 37.2267 & 0.16 & -58 06 31.780 	&-0.11 & 15.0 & 15.5\\
75 & 10 21 37.2287 & -0.41 & -58 07 12.980  	& 0.34 & 15.6 & 17.2\\
78 & 10 21 38.5253 & -0.21 & -58 06 16.790  	& 0.07 & 16.1 & 17.3\\
\enddata
\tablenotetext{1}{These are merely reference numbers, also used in Figure \ref{astro.image}. The USVO returned more than 80 stars, many of which did not fit our criterion, as discussed in Section \ref{sec:astrometry}.}
\tablenotetext{2}{The fitted positions of the star can be found by ddd.dd-residuals}
\tablenotetext{3}{The given positions of the stars from USNO, epoch 1980.274483.}
\end{deluxetable}

\begin{deluxetable}{lcccccccccc}
\rotate
\footnotesize
\tablecaption{Location and Intenisty of OH Line Centroids for Roberts~22\tablenotemark{1}\label{data}}
\tablewidth{0pt}
\tablehead{\colhead{Line}&\colhead{Channel}&\colhead{Velocity}&\colhead{Peak Int.}&{Unc.}&\colhead{Integral Int.}&\colhead{Unc.}&\colhead{RA J2000}&\colhead{Unc.}&\colhead{DEC}&\colhead{Unc.}\\
&\colhead{ATCA}&\colhead{km/s}&\colhead{Jy/Beam}&{Jy/Beam}&\colhead{Jy}&\colhead{Jy}&\colhead{h m s}&\colhead{s}&\colhead{\arcdeg~\arcmin~\arcsec}&\colhead{\arcsec}}

\startdata
1612&  60& -35.13& 2.1917& 1.02E-02& 2.2151& 1.78E-02& 10 21 33.884& .00182& -58 5 46.73& .01294\\
1612&  62& -34.41& 5.6757& 1.11E-02& 5.6536& 1.92E-02& 10 21 33.883& .00075& -58 5 46.74& .00545\\
1612&  64& -33.68& 9.1162& 1.34E-02& 9.0928& 2.31E-02& 10 21 33.876& .00055& -58 5 46.84& .00415\\
1612&  66& -32.95& 3.5755& 1.10E-02& 3.5849& 1.92E-02& 10 21 33.878& .00117& -58 5 46.90& .00874\\
1612&  68& -32.23& 3.2178& 1.02E-02& 3.2203& 1.78E-02& 10 21 33.877& .00122& -58 5 47.04& .00892\\
1612&  70& -31.50& 2.6848& 1.04E-02& 2.6622& 1.79E-02& 10 21 33.866& .00147& -58 5 47.03& .01086\\
1612&  72& -30.78& 1.4879& 1.00E-02& 1.4597& 1.71E-02& 10 21 33.856& .00251& -58 5 47.08& .01890\\
\enddata
\tablenotetext{1}{The complete version of this table is in the electronic edition of
the Journal.  The printed edition contains only a sample.}
\end{deluxetable}


\end{document}